\documentstyle[prl,epsfig,aps,multicol]{revtex}

\renewcommand{\narrowtext}{\begin{multicols}{2} \global\columnwidth20.5pc}
\renewcommand{\widetext}{\end{multicols} \global\columnwidth42.5pc}
\newtheorem{theo}{Theorem}
\newtheorem{lemma}{Lemma}
\newtheorem{cor}{Corollary}

\newtheorem{prop}{Proposition}

\newcommand{\proofend}{\raisebox{1.3mm}{
\fbox{\begin{minipage}[b][0cm][b]{0cm}\end{minipage}}}}
\newenvironment{proof}{{\noindent\it Proof:}
}{\mbox{}\hfill \proofend\\\mbox{}}

\newfont{\abg}{cmsl12} 

\def\nn{\nonumber \\}

\def\be{\begin{equation}}
\def\ee{\end{equation}}
\def\ba{\begin{eqnarray}}
\def\ea{\end{eqnarray}}
\def\pl{\label} 
\def\re{(\ref }
\def\rz#1 {(\ref{#1}) }   \def\ry#1 {(\ref{#1})}
\def\el#1 {\label{#1}\end{equation}}

\let\a=\alpha \let\b=\beta  \let\d=\delta
    
 \let\k=\kappa \let\l=\lambda \let\m=\mu
\let\n=\nu    
\let\t=\tau   
   
\let\O=\Omega \let\S=\Sigma

\def\0{\over} \def\1{\vec} \def\2{{1\over2}} \def\4{{1\over4}}
\def\5{\bar} \def\6{\partial}

\def\({\left(} \def\){\right)} \def\<{\langle} \def\>{\rangle}
 
\def\[{\lbrack} \def\]{\rbrack}

\def\wt{\widetilde}

  \def\CJ{{\cal J}}
  \def\CE{{\cal E}}       \def\CM{{\cal M}} 
\newcommand{\dR}{\mbox{{\sl I \hspace{-0.8em} R}}}

\begin{document}

\title{Second law of black hole mechanics for all 2d dilaton theories} 

\author{Norbert D\"uchting\footnote{e-mail: {\tt
norbertd@physik.rwth-aachen.de}}\\ \vspace{1mm}
{\em Institut f\"ur Theoretische Physik, RWTH Aachen,\\ D--52056
Aachen, Germany}\\ \vspace{3mm}
  and \\ \vspace{3mm}Thomas Strobl\footnote{e-mail: 
{\tt Thomas.Strobl@tpi.uni-jena.de}}\\ \vspace{1mm}{\em Institut f\"ur
Theoretische 
Physik,  FSU Jena\\
D--07743 Jena, Germany}
 \vspace{4mm} }

\date{September  2000}

\maketitle

{\tightenlines
\begin{abstract}
It is shown that all generalized two--dimensional dilaton theories with
arbitrary matter content (with a curvature independent coupling to
gravity) do not only obey a first law of black hole mechanics (which
follows from Wald's general considerations, if the entropy $S$ is
defined appropriately), but also a second law: $\delta S \ge 0$
provided only that the {\em null}\/ energy condition holds and that,
loosely speaking, for late times a stationary state is assumed.  Also any
two-dimensional $f(R)$--theory is covered. This
generalizes a previous proof of Frolov \cite{Frolov} to a much wider
class of theories. 

\vspace{0.3cm}



\vspace{0.3cm}
\end{abstract}
}

\narrowtext 
\renewcommand{\thefootnote}{\fnsymbol{footnote}}
\section{Introduction} 
Currently there is a noteworthy activity in finding a microscopic
statistical mechanical explanation for the entropy of black holes
(cf., e.g., \cite{strings,Ashtekar,Strominger2,Carlip}), where the most impressive
results where obtained within the string theoretical
approach. Underlying these attempts is the by now old analogy of black
hole mechanics (BHM) in classical Einstein gravity
\cite{Hawking1} with standard thermodynamics. This originally purely 
mathematical analogy was further supported by Bekenstein's Gedanken
experiments \cite{Beken} concerning the second law of thermodynamics
for a system in a spacetime with a black hole followed by Hawking's
observation \cite{Hawking2} that black holes should emit thermal
radiation when quantum effects are taken into account.

The Einstein-Hilbert action is just the (bosonic) leading order term in the
low energy effective action of string theory. Also other modifications
of Einstein gravity can be envisaged yielding agreement with current
experimental evidence. Thus the question arises \cite{TedMyers,Wald},
if, and in what sense, the three to four laws of BHM hold also in more
general theories than Einstein gravity.

At least what concerns the entropy of {\em stationary}\/ black holes,
Wald \cite{Wald} provided a very general definition of black hole
entropy $S$, which satisfies a (certain kind of) first law; in
particular, his definition of $S$ and the corresponding first law
applies to any diffeomorphism invariant theory of gravity in $D$
spacetime dimensions formulated in terms of a metric $g$. In general,
this entropy is no more proportional to the area of the
($D-2$--dimensional) black hole horizon, while it certainly reduces to
this result for the specific case of Einstein gravity. The upshot is
that, up to a proportionality constant, the entropy (at least for
stationary black holes) is defined as the integral over the black hole
horizon $\S$ of the (functional) derivative of the diffeomorphism
invariant action $L$ with respect to the curvature tensor $R_{\m \n
\rho \sigma}$ (having expressed all dependences of $L$ on derivatives
of $g$ in terms of the curvature of $g$ and its symmetrized covariant
derivatives; this is always possible as a consequence of the
diffeomorphism invariance of $L$ \cite{Wald}). Wald's first law of BHM
takes the form
\be
\frac{\k}{2\pi} \, \delta S = \delta \CE - \O_H^{(l)} \d \CJ_{(l)} \;
, \label{firstgeneral} \ee where $\k$ is the surface gravity of the
original (unperturbed) stationary spacetime, $\CE$ is the ``canonical
energy'', defined as the evaluation of the Hamiltonian (for Killing
time evolution of the original spacetime) on the classical solutions,
and, similarly, the $\CJ_{(l)}$'s, $l = 1, \ldots, r$, are ``canonical
angular momenta'' associated with $r$ axial Killing vectors
($\O_H^{(l)}$ being the appropriate angular velocities of the horizon;
certainly $r \le 1$ for spacetime dimension $D = 4$ and $D=3$, while
$r=0$ for $D=2$). For pure Einstein gravity, $\CE$ coincides with the
ADM mass, but, e.g., for Einstein-Maxwell theory it also yields the
standard contribution proportional to the change in the charge $Q$ of
the black hole \cite{Wald2}.

While there thus exists a formulation of the first law of BHM that applies
to sheer any conceivable theory of gravity, formulated even in
arbitrary spacetime dimension $D \ge 2$, the generality of the second
law for the so defined entropy is an essential and in general
very complicated open problem.

The second law of BHM states that during any dynamical process the
entropy $S$ can only increase, or, more precisely, \be \d S \ge 0
\quad \mbox{provided} \quad T_{\m\n} \, u^\m u^\n \ge 0 \; \; \forall
\, u^\m , \, u^\m u_\m \ge 0 \, .
\label{second} \ee
Here $T_{\m\n}$ is the energy momentum tensor of the matter content of
the theory and the second half of (\ref{second}) is known as the weak
energy condition.  This statement has been proven by Hawking
\cite{DeltaA} for Einstein gravity in four spacetime dimensions under
the (unproven) assumption that cosmic censorship holds true. The
proof, which recently has been put somewhat into question in
\cite{Nico} (but cf.\ also \cite{gall}), 
relies on the fact that in this case $S$ has the geometrical meaning
of (a quarter of) the area of the black hole horizon, and the weak
energy condition is brought in by the additional use of the specific
form of the Einstein field equations.  It is by no means clear, if
(and under what conditions) the entropy $S$ defined by Wald satisfies
a second law \re{second}) for more general gravitational
theories. However, for an analogy of BHM with thermodynamics, the
validity of a second law seems indispensable!

In a pioneering work, Frolov \cite{Frolov} showed that there is a
first {\em and\/} a second law of black hole mechanics for two--dimensional
string inspired dilaton gravity coupled to a complex scalar field (the
``tachyon'') and a $U(1)$ gauge field. The entropy $S$ defined by
Frolov coincides, furthermore, with the one defined by Wald.
In the present 
note, we want to show that
$S$ satisfies a second law for
 {\em arbitrary}\/ two--dimensional
dilaton theories with, moreover, arbitrary matter content (provided
the coupling to matter fields does not contain the curvature of the
metric explicitly). The second law we will find takes the strong
form\footnote{This formulation reproduces the result in the particular
case considered by Frolov. Note that the quantity $T_{\m\n}$ below
does {\em not}\/ coincide with what Frolov denoted by $T_{\m\n}$ and
that, on the other hand, the condition formulated by him may be
reexpressed as the condition below on the energy momentum tensor.}
\be \d S \ge 0 \quad \mbox{provided} \quad T_{\m\n} \, u^\m u^\n \ge 0
\; \; \forall \, u^\m , \, u^\m u_\m = 0 \, .
\label{secondprime} \ee
The latter requirement is only the {\em null}\/ energy
condition. Correspondingly 
Eq.\ \re{second}) follows as a trivial consequence from (its stronger
version) Eq.\ \re{secondprime}).  $T_{\m\n}$ is again the energy
momentum tensor, defined according to \be {T}_{\m\n} =
\frac{-4\pi}{\sqrt{-\det{g}}}\: \frac{\delta L_{matter}}{\delta g^{\m\n}}
\; , \label{T}\ee or, more generally, according to the formula
$T_{\m\n}  = -4\pi e_{a (\m} \, 
\(\d L_{matter}/\d e_{a}^{\n)}\) / \det ( e_\rho^b)$, if the matter
part of the action $L_{matter}$ depends also on a vielbein $e^a_\m$,
as is the case for fermionic fields (cf., e.g., ref.\  \cite{Heiko}
for  further details).  

Eq.\ \re{secondprime}) will be established under the assumption that
for late ``times'' $\dot S$ approaches zero sufficiently fast, 
i.e.\ there is a function $F$ within a certain class of allowed functions
(the class depending on the given action functional) such that
$\lim_{\t \to \infty} \dot S/F =0$, where $\t$ is a future directed
affine parameter along the (null) horizon and $S$ is fully defined in
terms of geometrical quantities defined on the horizon.  For a  more
precise formulation of this assumption  cf.\ Theorem
\ref{generaltheo} below, which is our main result. We remark that 
the assumption is, e.g., automatically satisfied, if timelike
infinity $i^+$ has a neighborhood which is vacuum (all matter fields
vanish there), cf.\ Corollary 2.

The organization of this note is as follows: In the next section we
recall the action and field equations for the class of
two--dimensional dilaton theories considered here and specify Wald's formula
for the entropy $S$ to these models. In the following section we then
review (in own words) Frolov's nice proof of the second law. At first
sight, his proof seems applicable only to a very restricted subclass
of models.  However, by a rather simple change of variables in the
action functional, this proof may be extended to the whole class of
theories.
With some patching and continuity considerations, global obstructions
to  the change
of variables will be shown to not spoil the final statement. The
significance of the result relies primarily in its {\em generality}\/
as well as in its close analogy with the (much more involved)
four--dimensional Einstein case.  

Instead of a generalized dilaton theory we may also take
$L_{grav}\propto\int d^2x\, \sqrt{-g}f(R)$
as for the gravitational part of the action, where $f$ is any nonlinear,
twice differentiable function of the curvature scalar $R$. The matter part
is again assumed to not depend  on $R$ explcitly. This is shown in a
separate  section. A final
section then  contains a brief summary 
and outlook on  possible further developments.

\section{The class of theories considered}
\label{sec:class} 
\subsection{Generalized two--dimensional dilaton theories}
In two spacetime dimensions the integrand of the Einstein-Hilbert
action $L_{EH}=\int d^2x\, \sqrt{-\det g}\, R$ becomes a total derivative
(locally) and thus does not yield any field equations. Introducing an
additional scalar field $\Phi$, the ``dilaton'', as a second
``gravitational'' variable beside the spacetime-metric $g_{\m\n}$, one
may consider the
following class of actions 
\cite{Banks} for 2d gravity theories as a substitute  of the
Einstein-Hilbert  action in higher
dimensions:
 \ba L_{grav}[g ,\Phi] &=& \frac{1}{4\pi} \int_{\CM} d^2x
\sqrt{-\det 
  g} \; [U(\Phi)R+ \nn && \qquad \qquad + V(\Phi) g^{\m\n} \6_{\m}\Phi
\6_{\n}\Phi +W(\Phi) ] \, .  \label{grav} \ea Here $R$ denotes the
Ricci scalar of the Levi--Civita connection of $g$ and $U$, $V$, and
$W$ are some essentially arbitrary functions (``potentials'') of the
dilaton, further specified below. Possible surface terms have been
omitted in \re{grav}). The total action results upon adding some
matter part, 
\be L_{tot}[g,\Phi,\psi] = L_{grav}[g,\Phi] +
L_{matter}[g,\Phi,\psi] \, , \label{Ltot} \ee 
where all the matter
fields collectively have been denoted by $\psi$.  For further
discussion of generalized dilaton theories cf., e.g., the review
articles  \cite{Strominger,sammel,habil} and citations therein.
We will assume in
the following without further mention that all the $g$--dependence of
$L_{matter}$ may be expressed without using the curvature scalar or
its derivatives, with eventual multiple covariant derivatives entering 
in symmetric combinations 
only. The $\Phi$--dependence of $L_{matter}$ is unrestricted, however.

To give an example: In the case of a complex scalar field $f$
and a $U(1)$ gauge field $A_\m$ (with its curvature $F=dA$), the
matter part of the action can take the form \ba L_{matter} &=&
\frac{1}{2\pi}\int 
d^2x \, \sqrt{-\det g} \; \[
-\frac{\alpha(\Phi)}{4} F^{\m \n} F_{\m \n} + \label{matterexample} \\ 
&& \!\!\!\!\!\! \!\!\!\!\!\! \!\!  -\, \frac{\beta(\Phi)}{2} \,
(\6^{\m}+iqA^\m)f^* (\6_{\m}-iqA_\m)f + \frac{\m^2(\Phi)}{2} f^*f \] \; ,
\nonumber \ea
where $\a$, $\b$, and $\m$ are $\Phi$--dependent coupling constants
and a ``mass'' parameter, respectively.

The interpretation of $g$ {\em and}\/ $\Phi$ as the basic
gravitational variables may be backed up further by the example of the
spherically symmetric sector of $D=4$ Einstein gravity (or likewise
for some other dimension $D \ge 3$): Implementing
\cite{spher} the spherical ansatz $ds^2_{D=4}= g_{\m\n}(x^\rho) \,
dx^\m dx^\n - \Phi^2(x^\rho)\: \[d\theta^2 + \sin^2 \theta \, d\varphi^2\]$,
$\, \m,\n,\rho \, \in \{0,1\}$, into the $D=4$ action yields a
two--dimensional action of the form \re{Ltot}) with an explicitly
$\Phi$--dependent part $L_{matter}$ and shows that in this case
$\Phi$ corresponds to a component of the metrical tensor in the higher
dimension. In general, $\Phi$ does not have such a clear
interpretation; nevertheless, it seems advisable also for the general
class of 2d toy models to treat $g$ and $\Phi$ on a similar footing and
to regard the action \re{grav}) as the 2d analogue of the 4d Einstein
action. It is also this point of view that leads to the simple and 
``physical'' condition in Eq.\ \re{secondprime}).

The model considered by Frolov \cite{Frolov} results from Eqs.\ 
\re{grav} - \ref{matterexample}) upon the choice \be U \equiv V
\equiv\a\equiv \b= 2\pi\exp \Phi \: , \; W = 2\pi\l \exp \Phi \: ,
\label{Frolov} \ee and $\m = const$. The gravitational part of this
action 
governs the s-wave modes of the low energy effective action for 
string  theory in four dimensions.   
If the matter part of the action is taken to be a minimally coupled
massless, real scalar field, the action is the well--known CGHS model
\cite{CGHS}, which may be used as a toy model for Hawking radiation
(cf.\ e.g.\ \cite{Strominger} for a review as well as \cite{Mikovic}
for a different approach to this issue within the same model). The
significance of the CGHS model resides in the fact that it may be
fully integrated, at least on the classical level. This is certainly
no more true for the general action $L_{tot}$ with arbitrary $U,V,W$;
it can be integrated only in relatively specific cases (cf.\
\cite{Heiko} for a general discussion and a collection of integrable
cases).

\subsection{Entropy for 2d black holes}
 
According to Wald the entropy $S$ of a (stationary) black hole is
determined by the functional derivative of the total action with
respect to the curvature of the metric integrated over the event
horizon at an instant of time. In $D=2$, taking the integral reduces
to mere evaluation at the given point on the horizon. The functional
derivative of $L_{tot}$ with respect to the curvature obviously yields
just a term proportional to $U(\Phi)$.

The explicit formula of Wald  \cite{Wald} 
prescribes the prefactor. 
Let us remark, however, that clearly the resulting entropy can be
rescaled if the {\em total\/} action of the theory under discussion is
multiplied by a prefactor. Note that the thus rescaled entropy still
satisfies an equation of the form \re{firstgeneral}) with the {\em
same\/} surface gravity $\kappa$, since the latter quantity is
determined by the geometry of the solutions, which clearly is not
changed by a different overall factor to the action; this is possible,
since then also $\CE$ and $\CJ$ will be rescaled by the same factor.
In the case of toy models with no underlying experiments it is hardly
possible to fix prefactors in front of the action, and, on the same
footing, having fixed the prefactor in the action, we think that one
should not regard the prefactor of the entropy to be fixed at the same
time. In the present case we decide for the simplest normalization,
\be S=  U(\Phi)\,|_{\, \mbox{event horizon}} \; ,\label{S} \ee 
having chosen the prefactors in \re{grav}) so that this formula agrees 
with what one obtains also by naive application of the formula provided in 
\cite{Wald}. (We have chosen conventions in which $R_{\mu\nu\rho}{}^\sigma$ 
is defined by $2[\nabla_\mu,\nabla_\nu]\omega_\rho=R_{\mu\nu\rho}{}^\sigma
\omega_\sigma$ on 1-forms $\omega_\mu$ and the norm of a spacelike vector 
is positive).  Still, one should keep in mind the eventual possibility
to reinterpret $a S + b$ for some constants $a$, $b$, as the ''true''
entropy.  This freedom may, e.g., be essential (although not
sufficient in general) to ensure positivity of the entropy. Note also
that adding a constant $b$ amounts just to the addition of a total
divergence to the Lagrangian \re{grav}).

The above result \re{S}) for the entropy
of a 2d black hole in a theory defined by the action $L_{tot}$ may be
obtained also by various other approaches (cf., e.g.,
\cite{Thermo,Frolov}).
The strength of Wald's approach is its generality and simplicity,
while in any known particular case it reproduces the result of other,
generically more limited approaches.

Strictly speaking, Wald's definition of the entropy applies only to
{\em stationary}\/ black holes. But this definition is extended in the
second paper of ref.\ \cite{Wald} to an entropy for nonstationary
black holes $S_{dyn}$ which happens to coincide with the original
prescription in the case of 2d dilaton gravity theories; at an
``instant of time'' the entropy $S$ is defined by $U(\Phi)$ evaluated
at the cross section of the event horizon with the spacelike
hypersurface specifying the instant of time. It is this notion of entropy
for which a second law will be established. 

\subsection{Technical Preliminaries}
\label{sec:technical}

To establish a second law, one needs the dynamics of the system under
consideration, i.e.\ one needs its field equations. In the present
context we will mainly need the variation of $L_{tot}$ with respect to
$g_{\m\n}$, which yields after some manipulations:

 \ba -\nabla_\m
\6_\n U(\Phi) + \2 g_{\m\n} \,
\left[ W(\Phi) - V(\Phi) \, (\nabla \Phi)^2\right] - && \nn + \; \6_\m
\Phi \6_\n \Phi \, V(\Phi) = T_{\m\n} - g_{\m\n} \, T \, , &\pl{eom}
\ea where $T_{\m\n}$ is the energy momentum tensor, defined in Eq.\ 
\re{T}), and $T$ is its trace. The variation with respect to $\Phi$ yields 
\be  U' \, R - V' \,  (\nabla \Phi)^2 - 2V \, \Box \Phi + W' =
\frac{-4\pi}{\sqrt{-\det{g}}}\: \frac{\delta L_{matter}}{\delta \Phi}
 \pl{eomPhi} \, . \ee It turns out that we will not need the remaining
field equations, $\delta L_{matter}/\d
\psi =0$.  As remarked already above, the combined field equations cannot be
solved in general; nevertheless, the validity of the second law, Eq.\
\re{secondprime}), still can be established, as we will demonstrate
in the subsequent section.

We now specify some technical assumptions on the allowed class of
potentials in the action, in part so as to exclude pathological cases,
but also adopting a rather pragmatic point of view here: We want to
present simple conditions on the potentials, still covering the
generic cases, at least from a physicist point of view.  (Although in
part possible, we do not intend to relax the conditions as much as
possible so that the main statements do still hold true, possibly in a
refined version).
Moreover, with respect to the restrictions on the potentials we focus
just on the functions $U$, $V$, and $W$ in \re{grav}); likewise
restrictions will apply to dilaton--dependent coupling constants such
as to the functions $\a$ and $\b$ in Eq.\ \re{matterexample}).

First, we require the potentials to be analytic\footnote{In most cases
  considered below this may be replaced by ``(sufficiently) smooth'',
  provided only eventual extrema of $U(\Phi)$ are isolated.}
functions on some {\em common}, connected, open domain of definition
$D \subseteq \dR$ and $U$ to be nonconstant. In this way we ensure,
e.g., that the action contains a kinetic term for the metric (and this
for all values of the dilaton).

Next, we restrict ourselves to such (generic) choices of the
potentials, for which eventual extrema of $U$ do not coincide with
zeros of $W$. E.g., in the case of higher order zeros of $W$,
inspection of the field equations shows that there would be solutions
with constant dilaton and vanishing matter fields for which the metric
remains completely unspecified. Clearly, such an action functional
would be no good toy model for Einstein gravity and should be excluded
from our considerations. We furthermore require for some technical
reasons that at any extremum $\Phi_0$ of $U$,
$\frac{V(\Phi_0)}{U''(\Phi_0)}\neq 2i,\;i\in{\rm I \!N}$ (cf.\
\cite{inprep}).

We focus on smooth ($C^\infty$) solutions $g$, $\Phi$, $\psi$ only and
always take spacetime $M$ to be a simply connected maximal extension
(cf.\ also \cite{KloeschII,KloeschIII}). Finally, we require that this
maximal extension of a local solution is determined solely by means of
the metric $g$; eventual values $\Phi_0 \in \partial D$ of $\Phi(x)$
for which one of the potentials diverges are supposed to lie on the
boundary $\partial M$ of spacetime 
{\em and}, provided such a boundary point exists, it is either
infinitely far away in terms of the spacetime metric $g$ (more
precisely, at least one geodesic approaching this point is complete)
or $R$ diverges there.

This includes an implicit restriction on the potentials. To illustrate
this, consider the simple action resulting from \re{grav}) upon the
choice $U:=\Phi$, $V:= 0$, and $W:=1$. Then the general, maximally
extended vacuum solution is easily seen to be Minkowski space with
$\Phi$ taking its limiting values $\pm \infty$ on the boundaries of
the diamond--shaped Penrose--Walker diagram. (Note that $\Phi(x)$ is
always nonconstant on $M$ and correspondingly only one of the Killing
vectors provides an isometry vector field of the total solution; which
one is determined by a (gauge--independent) integration constant
present in the solution for $\Phi(x)$.) Now let us replace $\Phi$ by
$\exp \Phi$ in the action; in the present case (for a more general
discussion cf.\ Eq.\ \re{Vtrafo}) below) this obviously changes only $U$
to $U=\exp \Phi$. By the above requirements, this choice of potentials
would not be permitted: The part of Minkowski spacetime which
previously was described by negative values of the dilaton is now
no more covered; the line in Minkowski space where now $\Phi \to -
\infty$ was previously just the internal line of vanishing dilaton field. 
Clearly at this line the metric is still well--behaved (even flat in
this simple example) and the line is by no means ``infinitely far away''. 

This example may be easily generalized. For any model with potentials
satisfying the above conditions, there are infinitely many, which do
not: Indeed, we only need to map any {\em proper\/} subset of $D$ onto
$\dR$ by a diffeomorphism $\Phi \to \widetilde \Phi(\Phi)$; the thus
induced potentials will violate the (reasonable) condition that the
maximal extension of the solutions is restricted solely by the metric.
So, ``artificial parametrizations'' of the dilaton field $\Phi$ are
excluded by the above conditions. 

Still within the allowed class of potentials there remains a freedom
to reparametrize the dilaton in a globally well--defined manner, which
we will make use of below. Note that while $U$ and $W$ transform like
ordinary functions under such diffeomorphisms $\wt \Phi: D \to I
\subset \dR$, $V$ is  density of weight two:

 \be U \mapsto
U \circ  f , \quad W \mapsto W \circ f , \quad
V \mapsto  f'^2 \cdot V \circ  f \; . 
\label{Vtrafo} \ee
Here $\circ$ denotes the composition of maps, prime differentiation
with respect to the argument, $\cdot$ pointwise multiplication, and
$f\equiv \wt\Phi^{-1}$ the inverse to the function $\wt \Phi$.

We conclude this section with some facts about the vacuum theories
(cf.\ \cite{KloeschII,inprep} for the details). First, for any vacuum
solution there is a Killing vector field, the integral lines of which
coincide with lines of constant dilaton. Thus, in the vacuum case the
horizon is a Killing horizon.  Under the assumptions specified above,
it may be shown \cite{inprep}, furthermore, that on the Killing
horizon the dilaton never assumes a critical value of $U$.

\section{Proof of the second law}
\subsection{Frolov's proof for the case given in Eq.\ \re{Frolov})}
\label{sec:Frolov}
By definition of the event horizon as the boundary of the causal past
of future null infinity, the event horizon is a null line. (The
cautious addition ``at its regular points'' in \cite{Frolov} is not
necessary, since in two spacetime dimensions there is no room for any
``cusps'' \cite{Nico} as in higher dimensions). Let $l^\m$ be a
generator of this line, such that 
\be l^\m \nabla_\m l^\n=0 \; .
\label{geo}\ee 
Such a choice of $l$ is always possible, since clearly
{\em any\/} null line in $D=2$ coincides with a geodesic. Next we
contract the field equations \re{eom}) with $l^\m l^\n$: 
\ba l^\m l^\n T_{\m\n}&=& - l^\m\6_\m \[ l^\n \6_\n U(\Phi)\] +
(l^\m \6_\m \Phi)^2 \, V(\Phi) \equiv \label{master}\\
&\equiv&\left(V(\Phi)-U''(\Phi)\right)\dot{\Phi}^2-U'(\Phi)\ddot{\Phi}
\, , \nonumber  \ea 
where use of Eq.\ \re{geo}) has been made. In the second line of this
equation a dot denotes differentiation with respect to the affine
parameter of $l$, which we will call $\tau$ in the following. In view
of Eq.\ \re{S}) and our goal to obtain a condition on the increase of
$S$, the right-hand side of the still general Eq.\ \re{master}) 
is somewhat demotivating.

In the case \re{Frolov}), considered by Frolov, now the following
simplification occurs in Eq.\  \re{master}):
Clearly to show that
$U|_H = 2\pi \exp{\Phi}|_H$ is increasing, $H$ denoting the event
horizon here and in what follows, it is sufficient to show that
$\Phi|_H$ itself is increasing.  Since, moreover, $U'' = V$ for the
choice \re{Frolov})
Eq.\ \re{master}) simplifies
to
\be \ddot \Phi = -\frac{1}{2\pi}\exp (-\Phi) l^\m l^\n T_{\m\n} \;
. \label{Frolovglg} \ee
Assuming that at late times the black hole settles down to a
stationary state, $\dot \Phi ( + \infty)=0$, Eq.\ \re{Frolovglg}) may
be integrated to yield \be \dot \Phi(\tau) =\frac{1}{2\pi}
\int_\t^\infty
\[\exp{(-\Phi)} l^\m l^\n T_{\m\n}\](\l) d \l \, . \ee
Since the right-hand side of this equation is positive upon the
 assumption $l^\m l^\n T_{\m\n} \ge 0$, the second law, Eq.\
\re{secondprime}), is established. 

There still is a small loophole in the above proof, if we require only 
$\lim_{\t \to \infty}\dot S =0$. Indeed, according to Eqs.\ \re{Frolov}) and 
\re{S}), the above asymptotic condition {\em implies\/} 
$\lim_{\t \to \infty}\dot \Phi =0$ {\em only if\/} $\Phi$ does not
approach $- \infty$ for $\t \to \infty$. At least for generic choices
of $L_{matter}$, {\em including\/} the case considered by Frolov (cf.\
Eq.\ \re{Frolov})) as well as
dilaton--independent and conformally invariant couplings to matter
fields, it may be shown by means of the field equations
that $\Phi \to - \infty$ implies $R \to \infty$. This to happen for
$\tau \to \infty$ is, however, in conflict with the notion of a
horizon and thus may be excluded in the present case. 

\subsection{The general case}

In the notation introduced above and making use of Eq.\ \re{S}), Eq.\
\re{master}) takes the form 
\be \ddot S = - \left[ l^\m l^\n T_{\m\n} - V(\Phi) \dot \Phi^2
\right]_H \; . \label{dotdotS} \ee

First of all, we observe that if $V(\Phi)$ is a nonpositive
function of $\Phi$\footnote{This is what one would expect for ordinary
scalar matter fields because of positivity of energy} or, more
generally, if  one can ensure that
evaluated on the event horizon this function is always nonpositive, we
can integrate \re{dotdotS}) directly to establish \re{secondprime}) 
under the late--time assumption 
\be \lim_{\t \to \infty} \dot S = 0 \; . \label{asymptot} \ee 

\begin{prop} If the potential $V$ in \re{grav}) is nonpositive and
  \re{asymptot}) holds true, the entropy-function $S$ obeys the
second law \re{secondprime}). \label{prop1}
\end{prop}

Note that for the string inspired dilaton theory considered in the
previous subsection as well as for a spherically reduced gravity
theory, the potential $V$ is {\em positive\/} definite, and the
assumption for the above reasoning is thus not satisfied in these
cases. On the other hand, many authors consider the action functional
\re{grav}) with $V \equiv 0$; then Proposition \ref{prop1} may be
applied. (This result should, however, not be confused with one for
the general theory \re{grav}); we will come back to this issue below).

The situation improves if one knows that $U$ is some strictly
monotonic function of its variable $\Phi$ evaluated along the horizon.
Let us assume first that $U$ is monotonically increasing (analogous
considerations apply to the decreasing case). Then we can make use of
the relation $\dot{S}>0\Leftrightarrow\dot{\Phi}>0$.  Employing Eq.\
\re{master}), on the other hand, we obtain the integral equality \be
\dot{\Phi}(\tau)=\int_\tau^{\infty}\frac{l^\mu l^\nu
  T_{\mu\nu}+\dot{\Phi}^2(U^{\prime\prime}-V)}{U^\prime}(\lambda)\,d\lambda
\label{huch}
\ee From this it is obvious that the second law is satisfied, if 
$U''-V$ is nonnegative, at least at the horizon.  In summary:

\begin{prop} If the potentials in \re{grav}) are such that 
$U$ is monotonic
and $V-U''$ is
nonpositive and if $\lim_{\t \to \infty} \dot \Phi = 0$ holds true,
the entropy-function $S$ obeys the second law \re{secondprime}).
\label{prop2}
\end{prop}
 
In string inspired dilaton gravity  $U$ is monotonic and 
the expression $(U''-V)$ vanishes identically; 
Proposition \ref{prop2} is the
straightforward generalization of the considerations in the previous
section \ref{sec:Frolov}.

Still, in general we will not be able to ensure that either $V|_H$ or
$[V-U'']_H$ in a given action functional is nonpositive and the
applicability of Propositions \ref{prop1} and \ref{prop2} seems to be
quite limited.  However, as noted already at the end of the previous
section, the potentials in the action functional change under a
reparametrization of the coordinate $\Phi$ (in a one-dimensional
target space), cf.\ Eq.\ \re{Vtrafo}). While obviously the sign of $V$
evaluated on the event horizon can not be changed by such a
transformation, the sign of $V-U''$ can, and by these means we will
now in fact be able to establish the second law under much more general
conditions. Defining the following nonnegative function
\be 
F_0(Z) := \exp\left[\int^Z\frac{V(z)}{U'(z)}\,dz\right] \; ,
\label{F} \ee 
where a constant of integration is assumed to be fixed somehow within
the integral, we have
\begin{prop}
\label{Umon}
  If $U$ in the action functional \re{grav}) as a function of the
  dilaton $\Phi$ has no extrema on its domain of definition $D$ and
  the late time assumption $\lim_{\t \to \infty} \dot S
  /F_0(\Phi(\tau))=0$ holds, then the entropy S, defined in Eq.\ 
  \re{S}), obeys the second law \re{secondprime}). \label{prop3}
\end{prop}

\begin{proof} 
  We can get rid of the term $U''-V$ in Eq.\ \re{master}) by an
  appropriate reparametrization $\Phi=f(\phi)$ of the dilaton. With
  \re{Vtrafo}) we see that $f$ has to obey \ba
  0&\stackrel{!}{=}&\tilde{U}''(\phi)-\tilde{V}(\phi) \equiv \nn
  &\equiv& \frac{d^2 f}{d\phi^2}U'(f(\phi))+\left(\frac{d
  f}{d\phi}\right)^2\left[U''(f(\phi))
  -V(f(\phi))\right]\;.\label{bed}\ea After one integration this may
  be verified to become
\be\left|U'(f(\phi))\frac{d
      f}{d\phi}\right|=F_0(f(\phi))\: .\label{U'f'}\ee

We choose $f$ to be orientation preserving, i.e.\ $f'(\phi)>0$.  Our
assumption $U'\neq 0$ and the regularity of $V$ within $D$ guarantee
that the derivative of $f$ with respect to $\phi$ does not vanish or
diverge.  $f$ is  a (globally well-defined)
diffeomorphism from some domain to the common domain of definition $D$
of the potentials.  Now our claim is easily established by noting that 
\be
\dot{S} = {\rm sgn}(U')\,F_0\,\dot{\phi} \label{hihi}
\ee
and that (using Eq.\ (\ref{huch}) for the redefined field $\phi$ and,
in a second step, the null energy condition)
\be {\rm
  sgn}(U')\,\dot{\phi}(\tau)=\int_\tau^\infty \frac{l^\mu l^\nu
  T_{\mu\nu}}{|U'(f(\phi))f'(\phi)|}(\lambda)\,d\lambda\;\ge 0 \; .
\label{proof} \ee 
\end{proof}

The condition in Proposition \ref{Umon} on the late time behavior is
implicitly one on the asymptotic behavior of $\Phi(\tau)$, since
$S(\tau) \equiv U(\Phi(\tau))$; here again $\tau$ is an affine
parameter along the event horizon. For any finite value of $\tau$, the
function $F_0$ is nonzero by its definition \re{F}) upon our conditions
on the potentials $U$ and $V$. However, $U'$ can vanish at the {\em
boundary\/} $\partial D$ of its domain of definition (which does not
belong to $D$ itself, being an open interval in $\dR$) and similarly
$V$ can diverge there. Thus in general the condition on the late time
behavior is different from Eq.\ \re{asymptot}); it is stronger
(weaker) whenever $\lim_{\Phi \to \partial D_\pm} F_0(\Phi)$ becomes
zero (divergent).  Whenever $G_0(\Phi) := \int^\Phi
\left(V(z)/U'(z)\right) dz$ does not approach {\em minus\/} infinity
at one of its boundary points, we consequently may also drop the
function $F_0$ in the condition of Proposition \ref{Umon}.

Note that in the vacuum theory always $\dot \Phi \equiv 0$ along the
horizon (cf.\ the remarks at the end of section \ref{sec:technical}).
Thus even in the case that $F_0(\Phi)$ vanishes on one of the two
boundary points $\partial D$, the above condition may be well
motivated by the requirement of a sufficiently fast approach of the
spacetime in the neighborhood of the horizon to the one of a vacuum
black hole.

Mostly in the literature the potential $U$ in the action functional
\re{grav}) is put to the identity map right at the beginning. We thus
specify the above Proposition to this particular case.

\begin{cor} For $U=\Phi$ and  $$\lim_{\t \to \infty} \exp 
  \left(-\int^{\Phi(\tau)} V(z) dz \right) \, \dot \Phi (\tau) = 0$$
  the second law \re{secondprime}) holds true.
\end{cor}

\smallskip

There is also another possibility for establishing Proposition
\ref{Umon}, which we intend to sketch as it provides an interesting
alternative point of view.  It makes use of a trick which is by now
well--known in the context of the general dilaton theories \re{grav}),
although in the present context one has to be somewhat careful as we
will emphasize below (in particular if we want to generalize the
considerations to nonmonotonic potentials $U$):  By a dilaton-dependent
conformal redefinition of the metric we can effectively {\em remove\/}
the potential $V$, which was potentially disturbing when proceeding
from Eq.\ \re{dotdotS}).  In the notation introduced above, this
transformation takes the form
\be g_{\mu\nu}\rightarrow \widetilde g_{\mu \nu} : = F_0(\Phi(x)) \,
g_{\mu\nu}\;,
\label{trafo}\ee
where the function $F_0$ is the same function as the one encountered
already above in quite a different context (reparametrization of the
dilaton, with $g$ held fixed). In terms of the rescaled variables the action
functional \re{grav}) reads \be L_{grav}[\wt
g,\Phi]=\frac{1}{4\pi}\int_{{\cal M}}d^2x\,\sqrt{-\det \wt
  g}\left[U(\Phi) \wt R
  +\frac{W(\Phi)}{F_0(\Phi)}\right]\;,\label{Aunphys}\ee where $\wt R$
is the curvature scalar of the metric $\wt g$.  Repeating the steps
leading to Eq.\ \re{dotdotS}), we obtain \be
\frac{d^2S}{d\wt\tau^2}(\wt{\tau})=-
\wt{l}^\mu \wt{l}^\nu T_{\mu\nu}\le 0\;,
\label{2hs}\ee where $\tilde{\tau}$ denotes the affine parameter of the
horizon in the auxiliary spacetime with metric $\wt g$, and
$\wt{l}$ is the corresponding generator satisfying
$\wt l^\m \wt \nabla_\m \wt l^\n=0$. Thus while $S$ will in general
not be a convex function of $\t$, cf.\ Eq.\ \re{dotdotS}),
as a function of $\wt \t$ it is. We then may proceed as for Proposition 
\ref{prop1} for $S(\wt \tau)$, taking care, however, of the 
relation between $\tau$ and $\wt \tau$. 

The affine parameter $\tilde{\tau}$ along the (lightlike) image of the
event horizon is connected to the respective affine parameter in the
physical spacetime with metric $g$ by (cf., e.g., \cite{waldbuch}) \be
\frac{d\tilde{\tau}}{d\tau}\propto F_0(\Phi(\tau))\;.\label{tt}\ee
Thus (by an appropriate choice of the proportionality constant) 
\be
\frac{dS}{d\tau}= F_0(\Phi(\tau)) \, \frac{dS}{d\tilde{\tau}}
\label{ent} \, , \ee
Thus from Eq.\ \re{2hs}) with $\lim d S/d\wt \tau = 0$ we reobtain
Proposition \ref{Umon}. 

Alternatively, but equivalently
to this second approach, is the use of a different generator of the 
event horizon  $\wt l$  than the one in
\re{geo}). Introducing  $\wt l$ according to 
$$  \wt l^\m \nabla_\m \wt l^\n = -\frac{V(\Phi)}{U'(\Phi)} 
\left(\wt l^\mu \partial_\mu \Phi\right) \wt l^\nu
\, , $$
the term including $V$ in Eq.\ \re{master}) cancels and, using $\wt
\tau$ as the flow parameter of $\wt l$, we again obtain \re{2hs}).

In the context of potentials with nonvanishing $U'$ both approaches,
the one with a redefined dilaton field $\phi$ and the one with a
conformally rescaled metric $\wt g$ (respectively a new parameter $\wt
\tau$) are equally applicable. However, when we turn to the case of the 
general class of potentials, restricted only by the conditions
presented in section \ref{sec:technical}, the former method turns out
to be much better suited. In fact, we managed to establish Theorem
\ref{generaltheo} below only by means of the first method. 

\smallskip

Before we turn to the case of nonmonotonic potentials $U$, however,
let us first unify the conditions in Propositions \ref{prop1},
\ref{prop2}, and \ref{prop3} (for the case of potentials with $U' \neq
0$ everywhere). As the propositions are formulated right now, the
respective fall-off conditions for late times are in part
different. E.g., if $U'' \equiv V$, then the function $F_0$ in Eq.\
\re{F}) becomes (proportional to) $U'$ and $\lim_{\t \to \infty} \dot
S /F_0(\Phi(\tau)) = \lim_{\t \to \infty} \dot \Phi$; but if, more
generally, $V-U''$ is just a nonpositive function, $\lim_{\t \to
\infty} \dot \Phi=0$ is a different condition than the one used in Proposition
\ref{prop3}, which, however, is applicable as well. 

To incorporate also Proposition \ref{prop2} as a particular case into
Proposition \ref{prop1} we need to relax the conditions on the
function $F_0$.  In fact, obviously in order to proceed along the
lines of the proof for Proposition \ref{prop2}, it is sufficient to
replace the left hand side of Eq.\ \re{bed}) by any nonnegative
function of $\phi$: Making for this function the $f$--dependent choice
$\left(f'(\phi)\right)^2 \, P(f(\phi))$ with some nonnegative function
$P$, we merely need to replace $F_0$ in Eq.\ \re{U'f'}) by its generalization 
\be 
F_P(Z) := \exp\left[\int^Z\frac{V(z)+P(z)}{U'(z)}\,dz\right] \; .
\label{Fneu} \ee 
All the remaining relations in the proof then remain the same (just
with $F_0$ replaced by $F_P$ and with now the correspondingly changed
function $f(\phi)$). Thus, in Proposition \ref{prop3} we are allowed
to replace $F_0$ by $F_P$ for {\em any\/} nonnegative function $P$. 

Now, if $U''-V$ is some nonnegative function $P(\Phi)$, we can choose
this function in Eq.\ \re{Fneu}) so as to obtain $F_P=|U'|$ ($f$ is
then seen to become just the identity map) and Proposition \ref{prop2}
becomes a particular case of Proposition \ref{prop3} (with $F_0$
generalized to $F_P$).  Similarly, if $V$ is some nonpositive
function, we can choose $P:= -V$ yielding $F_P\equiv 1$ and thus, in
the case of a theory with $U' \neq 0$, now also Proposition
\ref{prop1} becomes incorporated into (the above generalization of)
Proposition \ref{prop3}. 

The generalized assumption using $F_P$ instead of $F_0$ may be
obtained also in the second approach with a dilaton dependent
conformal transformation of the metric. Indeed, in repeating the steps
leading to Proposition \ref{prop1} with redefined metric $\wt g$ and
affine parameter $\wt \tau$, we need not cancel the potential $V$ in
\re{dotdotS}), but it obviously is sufficient to transform it to some 
nonpositive function $-P$. This, however, is achieved by just
replacing $F_0$ in Eq.\ \re{trafo}) by the function $F_P$ defined in
Eq.\ \re{Fneu}).

\medskip 

In the remainder of this section we want to extend our results to the
completely general case (no restrictions on the potentials except those 
mentioned in section \ref{sec:technical}) and show:
\begin{theo} Provided there is some nonnegative function $P: D \to \dR$ such 
that along the horizon $\lim_{\t \to \infty} \dot S
    /F_P(\Phi(\tau))=0$, with $F_P$ as defined in Eq.\ \re{Fneu}),
the entropy S as defined by Wald, cf.\ Eq.\ \re{S}), 
obeys the second law \re{secondprime}). \label{generaltheo}
\end{theo}
In particular we will now permit extrema of $U$ within its domain of
definition $D$. This implies that the function $F_P$ defined in Eq.\ 
\re{Fneu}) and evaluated along the event horizon can become divergent or
zero {\em within\/} spacetime. Either of the two approaches provided
to establish Proposition \ref{prop3} can then be applied only within
those sectors of the horizon where $\ln F_P =: G_P$ is
nondivergent. To prove Theorem \ref{generaltheo} we then need to glue
together different sectors. As mentioned already above, in this
context the first approach appears superior over the second one: In
the latter the parameter $\wt \tau$ becomes ill-defined at the
transition between two sectors and there one can thus hardly make
statements about $S$ as a function of $\wt \tau$. On the other hand,
by introducing an auxiliary dilaton field $\phi$ within each sector,
the differential equations remain formulated in terms of an everywhere
well-defined affine parameter $\tau$ and continuity arguments may be
applied.

In the following we will make use of a small lemma, the proof of which
is obvious:
\begin{lemma}
  Let $\varphi(\tau)$ and $G(\tau)$ be smooth functions on some open
  interval $(a,b) \subset \dR$, with possibly a divergent limiting
  value at $a$ or $b$, and $\dot S(\tau)$ be a smooth function on $[a,b]$
  with $\dot S(b) \ge 0$, satisfying
\be \varphi = \exp (-G) \dot S \, . \ee
  Then we may conclude from 
\be \dot \varphi \le 0 \quad \mbox{and}
  \quad \lim_{\tau \to b} \varphi \quad \mbox{nonnegative} \label{bed2} \ee 
that $\dot S \ge 0$ on all
  of $[a,b]$. 
\end{lemma}
In the above $a$ may be replaced also by $-\infty$ and likewise $b$ by
$\infty$. Furthermore, $\lim_{\tau \to b} \varphi$ need not necessarily
exist in the condition for the lemma; it is already sufficient that
there is an $\epsilon >0$ such that $\varphi(\tau) \ge 0$ for all
$\tau \in (b-\epsilon,b)$. We will interpret the respective condition in Eq.\ 
\re{bed2}) in this manner. 

\noindent {\em Proof\/} (of Theorem \ref{generaltheo}):  
Due to the analyticity required for $U$, eventual zeros of $U'$ are
always isolated. Correspondingly, if $G(\tau) := G_P(\Phi(\tau))
\equiv \ln F_P(\Phi(\tau))$ diverges not just at an isolated point on
the horizon but on a whole segment, then $\Phi$ will be constant along
this segment and therefore $\dot{S}$ will vanish there. $\dot
S(\tau)$, on the other hand, is a well--defined smooth function of
$\tau$, since $U$ and $\Phi$ are smooth and thus differentialbe by
assumption (and $\tau$ is just the flow parameter of a globally
well--defined vector field $l$). Thus, different sectors may be glued
together by continuity of $\dot S$. If the sectors are separated by
segments of diverging $G_P$, then the gluing is even simplified, being
achieved by $\dot S =0$ in this case. For notational simplicity we
thus restrict our attention to the case where different sectors are 
separated by points $\tau_i$ along the horizon on which $G$ diverges. 

Within each sector we can again perform the reparametrization
$\Phi=f(\phi)$ with $f$ defined implicitly through $|U'(f(\phi)) \,
f'(\phi)| =F_P(f(\phi))$ (cf.\ Eq.\ \re{U'f'})). Introducing the
function $\varphi:={\rm sgn}\left(U'(\phi)\right)\dot{\phi}$, Eq.\
\re{hihi}) and Eq.\ \re{proof}) (differentiated with respect to 
$\tau$ and using the null energy condition again) take the form
\be \dot{S}=F_P\, \varphi  \quad , \qquad \dot{\varphi} \, F_P
=-T_{\mu\nu}l^\mu l^\nu\le 0\; .\label{equi}\ee Since $F_P$ is
nonnegative in each sector by construction, we only need to make sure
that the second condition in Eq.\ \re{bed2}) is satisfied so as to
conclude from Lemma 1 above that $\dot S \ge 0$ within the respective
sector. 

Theorem \ref{generaltheo} can now be proven by induction. Let
$\tau_1>\tau_2>\ldots$ denote the values of the affine parameter along
the horizon where $G$ happens to diverge. By our assumption we
know that $\varphi\to 0$ for $\tau\to \infty$. Thus in the rightmost sector, 
$\tau\ge \tau_1$, $\dot{S}\ge 0$ is established. 

For the induction let $\dot{S}\ge 0$ for $\tau\ge \tau_i$. The
continuity of $\dot{S}$ and the positivity of $F_P=\exp G$ assures
that $\lim_{\tau\to\tau_{i-}}\varphi$ is nonnegative. With
Lemma 1 we obtain $\dot{S}\ge 0$ for $\tau_{i+1}\le\tau\le\tau_i$ and,
by induction, thus for all $\tau$. 

\hfill \proofend

The condition on the late time behavior of the entropy of spacetime
may be dropped alltogether, if for late times spacetime becomes equal
to the one of a vacuum black hole: 

\begin{cor} \label{cor2}
If for all $\tau > \tau_0$ for some $\tau_0$  the event horizon of the
two-dimensional spacetime has a neighborhood which is vacuum
($T_{\m\n}=0$ in this region), then the second law \re{secondprime})
is satisfied for all times.
\end{cor}

This follows immediately from Theorem \ref{generaltheo} and section
\ref{sec:technical} (note in particular that according to  section
\ref{sec:technical} $\ln F_0$ is nondivergent for $\tau > \tau_0$).

\section{Second law for  $\lowercase{f}(R)$-theories}
Instead of taking recourse to a dilaton field $\Phi$ so as to define a
nontrivial gravitational part of the action in two spacetime
dimensions, we may also consider higher derivative theories, where the
Ricci scalar $R$ in the Einstein action is replaced by a nonlinear
function of it. This leads us to consider (cf.\ also \cite{f(R),habil})
\be L_{geom}=\frac{1}{4\pi}\int_{\cal
M}d^2x\,\sqrt{-g}\,f(R)\label{geom}\ee 
 as for the
gravitational part of the action instead of
\re{grav}).  Correspondingly, the matter part of
the  action  is
then  also assumed not to depend on a dilaton, or alternatively, the dilaton
is understood as yet just
another scalar  scalar field among other matter fields  described
collectively 
by means of $\psi$ in $L_{matter}[g,\psi]$. Again $L_{matter}$ is
assumed to not depend explicitly on $R$. Then according to Wald the
entropy $S$ of a black hole in an $f(R)$-theory is nothing but $f'(R)$
evaluated on the horizon at a given instant of time (parametrized
by  the affine parameter $\tau$).

To have a nontrivial kinetic term for the metric, we should require
that the function $f$ is locally nonlinear, i.e.\ that there is no
interval (within the domain of definition of $f$) in which it is
linear. The physical requirement of nonlinearity may also be dropped
on a formal level, however, since a linear function $f$ yields a
constant entropy according to Wald, which then clearly satisfies 
a second law, too.  Finally, we require $f$ to be three times
differentiable (cf.\ the field equation Eq.\ \re{eomg}) below).

For {\em convex\/} functions
$f$, the model of the present section may be
considered  a special case of the one in the previous
sections. Indeed, choosing $U=\Phi$, $V=0$, and $-W$ as the Legendre
transform of $f$ within the action \re{grav}), and eliminating the
dilaton $\Phi$, we just reproduce the action \re{geom}). However,
even in the case of a general, nonconvex choice for the function
$f$ in the gravitational part of the action, we can establish a second
law. We thus will not take recourse to our previous results (which
then would have to be ``glued together'' appropriately along regions
in spacetime where $R$ takes a value for which $f'$ becomes extremal),
but rather show directly:

\begin{theo} If the spacetime in an $f(R)$-theory satisfies
$\lim_{\tau \to \infty} \dot S = 0$ or, more generally, if there exists
some strictly monotonically increasing  function $H$ with nonpositive
second derivative such that
$\lim_{\tau\to\infty}\dot{H}(S)=0$,   then the second law
\re{secondprime})  holds, i.e.\
 provided the matter fields satisfy the null energy condition, 
$S(\tau)$ can at most increase in time.  \label{2}\end{theo}

\begin{proof} The equations of motion of the class of
models under consideration are given by
\be -\nabla_\mu\nabla_\nu
f'(R)+\frac{1}{2}g_{\mu\nu}\left(f(R)-Rf'(R)\right)=T_{\mu\nu}-g_{\mu\nu}T
\label{eomg}\;. \ee
By contracting this equation with the generator $l$ of the horizon,
we  obtain
\be T_{\mu\nu}l^\mu l^\nu=-l^\mu\partial_\mu\left(l^\nu\partial_\nu
f'(R)\right)=-\ddot{S}(\tau)\;,\label{contr}\ee
where this equation is assumed to be evaluated on the horizon. By
integrating Eq.\ \re{contr}) Theorem  \ref{2}
is now established as in the previous section provided the late-time
assumption $\lim_{\tau \to
\infty} \dot S = 0$ holds.
  
Let us now consider the more general case
of the late-time assumption $\lim_{\tau\to\infty}\dot{H}(S)=0$. 
By introducing the function $\varphi=H(S)$ we easily obtain the relations
\be\dot{S}(\tau)=\frac{\dot{\varphi}(\tau)}{H'(S(\varphi(\tau)))}
\label{spunkt}\ee 
as well as
\be \ddot{S}(\tau)=\frac{\ddot{\varphi}(\tau)}{H'(S(\varphi(\tau)))}-
\frac{H''(S(\varphi(\tau)))\dot{\varphi}^2(\tau)}{H'^3(S(\varphi(\tau)))}
\;.\label{s2punkt}\ee
{}From Eq.\ \re{spunkt}) and our assumptions concerning $H$ we see that
nonnegativity of $\dot{S}$ is equivalent to nonnegativity of
$\dot{\varphi}$. Eq. \re{s2punkt}), on the other hand, can be
integrated for $\dot{\varphi}$ using \re{contr}) and our late-time
assumption  to yield
\be \dot{\varphi}(\tau)=\int_\tau^\infty d\lambda\,
\left(T_{\mu\nu}l^\mu l^\nu
H'(S(\varphi))-\frac{H''(S(\varphi))\dot{\varphi}^2}{H'^2(S(\varphi))}
\right)(\lambda)\;.\label{int}\ee
Using the positivity of $H'$, the
nonpositivity of $H''$ and the null energy condition, we conclude from
this equation that
$\dot{\varphi}$ is nonnegative. This is 
 equivalent to the proposition of the theorem. 
\end{proof}

As mentioned above, in a region where the function $f$ in the action
is convex  the geometric action may  be reformulated in terms of a
dilaton theory. By comparing the notations in the present section and
in the previous one, it is easy to see that the function $H'(S)$ is
the analogue of $1/F_P$. This function is strictly positive
and monotonocally decreasing. These are exactly the requirements we
posed on $H'$ in Theorem 2.    

One may ask whether the version of the theorem with the late-time
assumption involving the function $H$  is really a generalization of
the statement using the assumption $\lim_{\tau \to \infty} \dot S =
0$. As can be seen from Eq.\ \re{spunkt}), this is  the case only if
$H'(S)$ tends to zero and $\dot{S}$ diverges for large $\tau$  (without
$R$ diverging at the same time since this would be in conflict with
the notion of a horizon).
This, on the other hand,
can  happen only if the function $f$ in the action \re{geom}) is
defined on  an interval with a divergent limit at a boundary 
value (such as e.g.\ $f(R) = \tan R$). For functions $f$ which are
continuous on all of $\dR$ this part of Theorem \ref{2} may be dropped
without loss of generality.

\section{Conclusion and Outlook}
Our results are contained in the Theorems \ref{generaltheo} and
\ref{2} as well as in Corollary
\ref{cor2}.  We established the validity of the second law of black
hole mechanics for all generalized 2d dilaton gravity theories. The
class of theories was recapitulated in section
\ref{sec:class}: The gravitational part $L_{grav}$ of the action replacing the 
standard Einstein term in four dimensions has the form \re{grav})
(satisfying conditions on the potentials such as analyticity as
specified in detail in section \ref{sec:technical}), while the matter
part $L_{matter}$ was left essentially arbitrary. For the notion of an
entropy $S$ of a 2d black hole we applied the general formalism of Wald
\cite{Wald}, extrapolated to the dynamical situation, which yielded 
$U(\Phi)$ evaluated at the event horizon at a given instant of time.
(Here enters the only restriction on $L_{matter}$: it should not
depend explictly on $R$, since otherwise the formula for $S$ would
change).  We then showed that  during any dynamical process which
finally settles down (sufficiently fast) to a stationary (e.g.\
vacuum) situation the entropy $S$ can at most increase. Here we even
need not require the weak energy condition (or assume anything like
cosmic censorship); it is sufficient that the energy momentum tensor
of matter satisfies the null energy condition. 
We also showed the second law for models in which the gravity part of
the action is replaced by the geometrical one \re{geom}). For these
theories the entropy turned out to be $f'(R)$  evaluated on the
horizon.

We remark here in parenthesis that the s-wave sector (i.e.\ the
spherically symmetric sector) of Einstein gravity in $D \ge 3$
dimensions with arbitrary minimally coupled matter content is
contained as a particular example to the present considerations.  
The type of theories considered here may be of interest also as the
s-wave sector of more complicated dilaton gravity theories in higher
dimensions, e.g.\ as they arise generically in (super) string theory 
at low energies.

It may be worthwhile to extend our results to even more general 2d
gravity theories, such as theories with dynmical torsion:
\be L=\frac{1}{4\pi}\int_{\cal M} d^2x\,\sqrt{-g}\,
F(R,\tau^a\tau_a)\label{tor}\ee where $F$ is a sufficiently smooth
function and $\tau^a$ denotes the Hodge dual of the torsion 2-form
$De^a$. For functions $F$ with a well-defined Legendre transform there
is a formulation of these theories in terms of Poisson Sigma Models
(cf., e.g., \cite{habil}), which, on the other hand, also cover the
$f(R)$-theories and the general dilaton theories with monotonic
potential $U$ discussed within the present paper. It is thus plausible that a
second law may be established also for the class of theories where the 
gravity part of the action has the form of Eq.\ \re{tor}).

Concerning the first law of black hole mechanics we relied on the
general formalism of Wald guaranteeing the existence of such a
law. Still it may be worthwhile, also in combination with the present
second law, to discuss the resulting ``2d black hole mechanics'' in
further detail and from different physical perspectives. 

Our considerations remained on a purely classical and mathematically
rigorous level. Still, it would be interesting, if and in what sense
the results may be extended also into at least a semiclassical
regime. Is there, e.g., a ``generalized second law'' of black hole
mechanics when Hawking radiation and some approximate back reaction
are taken into account? For specific 2d models such issues have been
discussed already partially in the literature \cite{RSTetcdenkeich}. 

We finally come back to the issue of a microscopic explanation for the
entropy of black holes, which certainly also arises within the
two-dimensional setting. Here we mention two approaches which turned
out succesful in the case of 2+1 Einstein gravity with a cosmological
term, giving rise to BT(H)Z black holes \cite{BTHZ}. First, in the
spirit of the AdS/CFT correspondence  \cite{Maldacena},
Strominger \cite{Strominger2} obtained the statistical entropy  by
counting the degeneracies of the representations of the algebra of 
asymptotic symmetries of this spacetime. In the two-dimsional setting
this  idea was applied \cite{CadoniStrominger}
to a very particular theory of the type \re{grav}),
namely the so-called Jackiw--Teitelboim (JT)
model \cite{JTinklusiveRussen-siehebeimir}, which results 
upon the choice $U:=\Phi =:W$, $V:=0$, for the
potentials.

Another approach \cite{Carlip}, which turned out successful in the
case of 
BTHZ black holes, ascribes the black hole entropy to
tracing out the degrees of freedom of an effective theory living on
the event horizon. In two spacetime dimensions this corresponds to 
a quantum theory of point particles. Adapting Carlip's approach to
the general class of theories with action \re{grav}) \cite{Kunst}
did not reproduce the expected answer: In some cases it just produced
an ill-defined infinity, in other cases (such as the above mentioned
JT model) it produced the logarithm of the expected
result. Let us remark here, however, that by means of some additonal
steps one can obtain the expected result, at least for cases such as
the JT model or spherically reduced gravity and up to
prefactors. Roughly speaking this works as follows:\footnote{T.S.\ is
grateful for discussions with H.\ Verlinde and G.\ Kunstatter in this
context.} The point particle phase space found in \cite{Kunst}
was a copy of the two-dimensional
spacetime manifold, equipped with a specific symplectic form $\Omega$.
Now one keeps only the part of phase space corresponding to the
interior of the
black hole, appropriately euclideanizes $\Omega$, and then performs 
a second quantization with fermionic statistics; then $S$ may be 
identified with the logarithm of the dimension of the resulting
Hilbert space.

But even if correct numbers may be produced in such or another 
manner, there always remains the question for the underlying general 
principle and, all the  more, a principle which would work in general
dimensions (in analogy to Wald's general formalism for obtaining a
classical prediction for the entropy $S$, satisfying a first
law). The relatively large class of 2d gravity models discussed in
this note may give us decisive hints to the right answers in this
context.

The existence of a second law of black hole mechanics, despite still
on a classical level, is already an important step in this direction.
Combined with the first law and due to its close formal analogy to the
much more involved four dimensional situation, it should provide enough
motivation for further studies of such theories.

\begin{acknowledgements} T.S.\ is grateful to Valery Frolov for
  discussions on \cite{Frolov}. The work of N.D.\ is supported by the
DFG under the contract KA-246/16-1 and by the Graduiertenkolleg
``Starke und elektroschwache Wechselwirkungen bei hohen Energien''.
 \end{acknowledgements}

\widetext

\end{document}